\documentclass{article}

\pdfoutput=1

\usepackage{arxiv}

\usepackage[utf8]{inputenc} % allow utf-8 input
\usepackage[T1]{fontenc}    % use 8-bit T1 fonts
\usepackage{hyperref}       % hyperlinks
\usepackage{url}            % simple URL typesetting
\usepackage{booktabs}       % professional-quality tables
\usepackage{amsfonts}       % blackboard math symbols
\usepackage{nicefrac}       % compact symbols for 1/2, etc.
\usepackage{microtype}      % microtypography
\usepackage{lipsum}
\usepackage{mathrsfs,amsmath}
\usepackage{amssymb}
\usepackage{moreverb}
\usepackage{graphicx}
\usepackage{verbatim}

\newcommand{\cD}{{\cal{D}}}

\newcommand{\cF}{{\cal{F}}}

\newcommand{\bld}[1]{\mbox{\boldmath $#1$}}

\newcommand{\bbc}{\bld c}

\newcommand{\bbk}{\bld k}

\newcommand{\bbs}{\bld s}

\newcommand{\bbu}{\bld u}
\newcommand{\bbv}{\bld v}

\newcommand{\bbx}{\bld x}

\newcommand{\bbz}{\bld z}

\newcommand{\bbB}{\bld B}

\newcommand{\bbI}{\bld I}

\newcommand{\bbK}{\bld K}

\newcommand{\bbR}{\bld R}

\newcommand{\bbX}{\bld X}
\newcommand{\bbY}{\bld Y}
\newcommand{\bbZ}{\bld Z}

\newcommand{\bomega}{\bld \omega}

\newcommand{\bmu}{\bld \mu}
\newcommand{\bbmu}{\bld \mu}
\newcommand{\bbeta}{\bld \beta}

\newcommand{\beps}{\bld \epsilon}

\title{Rapid Numerical Approximation Method for Integrated Covariance Functions Over Irregular Data Regions}

\author{
  Peter Simonson\thanks{Corresponding author.} \\
  Department of Applied Mathematics and Statistics\\
  Colorado School of Mines\\
  Golden, CO 80401, USA \\
  \texttt{petersimonson@mines.edu} \\
   \And
  Douglas Nychka \\
  Department of Applied Mathematics and Statistics\\
  Colorado School of Mines\\
  Golden, CO 80401, USA
  \And
  Soutir Bandyopadhyay \\
  Department of Applied Mathematics and Statistics\\
  Colorado School of Mines\\
  Golden, CO 80401, USA
}

\begin{document}
\maketitle

\begin{abstract}
In many practical applications, spatial data are often collected at areal levels (i.e., block data) and the inferences and predictions about the variable at points or blocks different from those at which it has been observed typically depend on integrals of the underlying continuous spatial process. In this paper we describe a method based on \textit{Fourier transform} by which multiple integrals of covariance functions over irregular data regions may be numerically approximated with the same level of accuracy to traditional methods, but at a greatly reduced computational expense.
\end{abstract}

% keywords can be removed

\section{Introduction}\label{sec:intro}
\subsection{Motivation}\label{subsec:motive}

Due to advances in science and technology, spatial data from remotely sensed observations, surveys, and censuses are being gathered at a rapid pace (cf., \cite{Brad:2015a, Brad:2015b, Brad:2016, Nguyen:2012, Nguyen:2014}) and subsequently, this has created the opportunity to quantify spatial dependence and make predictions for many different kinds of processes and variables. Often these data are in the form of spatial averages over irregular and possibly overlapping regions. This feature makes it difficult to apply standard methods of spatial analysis which are typically built for point-referenced data. These kind of spatial problems are termed as \emph{change of support} problems (see \cite{Cressie:1993}) and our primary concern is to make statistical inferences about the values of a variable at spatial scales different from those at which it has been observed (cf.,~\cite{Gotway:2002}). For example, in the context of remote sensing and continuous geophysical variables, satellite data products can be reported as averages over a set of regions (e.g., intervals, areas, or volumes), but the interest is in spatial field varying over a continuum. When the underlying spatial field can be represented as a Gaussian process and  the observational data are linear functionals of the field, a standard statistical framework can be applied to make inferences about the spatial process. This approach depends on evaluating the covariance matrix among the observations. In the case of linear functionals being integrals over spatial regions, multi-dimensional integrals involving the covariance function may not have a closed analytical form and so a numerical method is required for such computations.  Efficient numerical methods in the literature are typically either tied to particular region geometries, require user intervention to allocate quadrature points, or achieve efficiency at a cost in accuracy (see \cite{Journel:2003, Martin:1994}). This work was motivated by the lack of  accurate numerical strategies to handle large change of support problems.  We present a new method of numerically approximating covariance integrals over irregular regions when the underlying covariance model is assumed to be (second-order) stationary. This method is efficient because it uses the discrete Fourier transform (DFT) and so can handle a large number of quadrature points to gain accuracy. In addition, our approximation is based on a coherent representation of the spatial process and so has a useful interpretation.  Finally, the proposed method deals with a single discretized process and discretized integrals for which the observation covariances and subsequent statistical inferences are exact. 

\subsection{Model}\label{subsec:model}

Consider a random field  $Y(\bbs)$  with mean function $\mu_Y(\bbs)$ and covariance function $c_Y(\bbs,\bbs^{\prime})$, for $\bbs,\bbs^{\prime} \in \bbR^d$. We also assume $Y(\bbs)$ to be defined over a domain $\cD\subseteq \bbR^d$.  Let   $B_1,\ldots,B_n$ be regions in $\cD$ and we have observation functionals given by $\bbz_i =\frac{1}{|B_i|} \int_{B_i} Y(\bbs)d\bbs$, where $|B_i|=\int_{B_{i}}d\bbs$ for $i=1,\ldots,n$. If $Y(\bbs)$ is a Gaussian process, then $\bbz$ follows a multivariate normal distribution (\cite{Song:2008}, \cite{Cressie:1993}). To simplify the exposition in this paper we only focus on two-dimensional spatial domains and with polygons serving as the regions of interest, but the proposed method can operate in arbitrary dimensions. Moreover, the regions of interest need not be polygonal, only well-approximated by indicator functions discretized over regular grids. See Section~\ref{algo} for details.\\
\vskip-1em
\noindent Define the mean of the observation functionals as
\begin{eqnarray}
 \bld \mu_i = E(\bbz_i)  =  \frac{1}{|B_i|} \int_{B_i} \mu_Y(\bbs)d\bbs,\ \mbox{for}\ i=1,\ldots,n. 
\label{eq:target0}
\end{eqnarray}

\noindent and the covariance matrix, $\bbK$, as

\begin{eqnarray}
\bbK_{i,j}=  cov(\bbz_i,\bbz_j)= \frac{1}{|B_i|} \frac{1}{|B_j|} \int_{B_i} \int_{B_j} c_Y(\bbu,\bbv) d\bbv d\bbu\ \mbox{for}\ i,j=1,\ldots,n. \label{eq:target}
\end{eqnarray}
Note that, in Eqn.~(\ref{eq:target}), the integrand is bounded and therefore, the interchange of expectation and integration is permissible. %using Fubini's theorem. 
In general, one might also consider an additional weight function in the integrand but we will omit this extension to simplify exposition. \\

\vskip-1em
\noindent Generally, two basic components of a spatial data analysis are prediction of the spatial field to locations that are not observed and estimating statistical parameters in the covariance functions and the observational model.  Here we emphasize the aspects of these operations that are challenging with change of support data.  
Typically, one includes an additional measurement error component, so the complete  observational model  is 
\begin{equation}
\bbZ_i = \bbz_i + \beps_i,
\end{equation}
with  $\beps$ being mean zero, Gaussian white noise with variance $ \tau^2 $ and independent of the process $Y$.
Under the assumption that $\bmu$, $\mu_Y(\bbs)$,  $\tau^2$ and any additional parameters in $c_Y$ are known, we have the standard Kriging prediction for 
$Y(\bbs)$ given as 
\begin{equation}
 \label{eq:target00}  
\widehat{Y}(\bbs) =  \mu_Y(\bbs) +  \bbk(\bbs)^T ( \bbK + \tau^2 \bbI ) ^{-1}(\bbZ - \bmu) 
\end{equation}

\vskip-1em
\noindent where,
 \begin{equation}
 \label{eq:target1}  
 \bbk(\bbs)_i = COV(Y(\bbs), \bbZ_i)=  \frac{1}{|B_i|}\int_{B_i} c_Y(\bbs,\bbu) d\bbu, 
 \end{equation} 
and $\bbI$ is the identity matrix. Thus we see that prediction will involve evaluation of $\bbK$ and also an additional integral for every prediction location. Our proposed method will give accurate approximations to this vector and makes it feasible for spatial prediction on a large and dense grid of locations. \\

\vskip-1em
\noindent Another important aspect of a spatial analysis is  estimating unknown statistical parameters. Following a maximum likelihood approach or nested within a Bayesian model, one requires evaluation of the negative log likelihood 
\begin{equation}
\label{nloglike}
-\ell( \bbZ) =\frac{1}{2} (\bbZ - \bmu)^T ( \bbK + \tau^2 \bbI ) ^{-1}(\bbZ - \bmu) +  \frac{1}{2} \log | \bbK + \tau^2 \bbI  |  +C,
\end{equation}
where $C$ is a constant and where both $\bmu$ and $\bbK$ may depend on other statistical parameters. From Eqn.~(\ref{nloglike}) we see that mean and covariance parameters for the underlying process can be deduced through the observational covariance matrix. However, maximization of this likelihood will  require re-evaluating  the integral expressions for $\bbK$ and $\bmu$ for any  parameters that enter Eqns.~(\ref{eq:target0}) or   (\ref{eq:target}) in a nonlinear manner. Our proposed method significantly reduces the computational burden associated with evaluation of $\bbK$ and $\bmu$ and so makes maximum likelihood methods, or related Bayesian inference feasible.

\section{Algorithms }
\label{algo}
In this section we review some of the previous work and describe our new method. Our new approach exploits the computational efficiency of the DFT and we refer to this as Fourier Approximations of Integrals over Regions (FAIR).  For certain families of covariance functions and regular shapes, such as rectangles and triangles, it is possible to obtain closed form expressions for the covariance matrix $\bbK$. However, a core set of spatial statistical applications typically involve at least two dimensions and covariance functions based on radial distances. Therefore, a closed form expression for $\bbK$ often does not exist. For instance, for the widely used Mat{\`e}rn family the closed form expressions are not available, even for rectangular regions.

\subsection{Direct quadrature}\label{subsec:dq}

A direct approach to approximate the key integrals given in Eqns. (\ref{eq:target0}),~(\ref{eq:target}) and~(\ref{eq:target1}) deals with a sum, up to a scaling factor, over a grid of locations restricted to the regions, i.e., a discrete, Riemann approximation to the continuous integral. 
In particular, suppose %${\bbs_k^G}$% 
$G$ be a regular grid of locations that include the spatial domain and let $\bbs_i^G = \{\bbs: \bbs\in B_{i}\cap G\},\ i=1,\ldots,n$, with $L_i^G$ the number of points in $\bbs_i^G$. Then 
$\bbK_{i,j}$ can be approximated  by  
 \begin{eqnarray}
%\frac{1}{|B_i|} \frac{1}{|B_j|} \sum_ {k,l}  I_{B_i}( \bbs_k^G) I_{B_j}( \bbs_l^G) c_Y(\bbs_k^G, \bbs_l^G) \boldsymbol{\Delta} \label{eqRieman}
%\frac{1}{\cC(\bbs_i^G)} \frac{1}{\cC(\bbs_j^G)} \sum_ {\bbs_{k}\in \bbs_i^G,\bbs_{\ell}\in \bbs_j^G}  c_Y(\bbs_k, \bbs_\ell) \boldsymbol{\Delta}, \label{eqRieman}
\frac{1}{L_i^G L_j^G}\sum_ {\bbs_{k}\in \bbs_i^G,\bbs_{\ell}\in \bbs_j^G}  c_Y(\bbs_k, \bbs_\ell)  \label{eqRieman}
 \end{eqnarray}
%where, $\cC(A)$ denotes the cardinality of a set $A$, $\boldsymbol{\Delta}$ represents a scaling term based on the grid resolution of the quadrature points (e.g. in a two-dimensional case $\boldsymbol{\Delta}=(\Delta x)^2 (\Delta y)^2$ ).  %$I_B(\bbs)$ are indicator functions over the observation regions. 
This approximation has the advantage that the discretized process on the grid is the stochastic basis for the statistical model. In fact the analysis will be exact if one replaces the integral expressions for the observation functionals by discrete sums implied by this Riemann approximation. 
 Although straightforward as a formula with a useful interpretation,  many authors have pointed out the computational burden in evaluating this double sum over a multidimensional grid.  Unfortunately for most problems and in particular for large spatial data sets the computations are prohibitive with single threaded codes. \\
\vskip-1em
\noindent To overcome the computational challenges inherent in Eqn.~(\ref{eqRieman}), \cite{Journel:2003} (henceforth [JH])  propose a Riemann approximation  described as ``centered regular discrete approximation with uniform weighting" that is particularly well-suited for semi-variogram (or covariance) approximation when the regions of interest are parallelograms or parallelepipeds. The key idea is to use many small, distinct grids tailored to each region. 
%More precisely, given two grids $G_{1}$ and $G_{2}$ such that $\cD\subseteq G_{i},i=1,2$, if $\bbs_{i}^{G_{1}} \subseteq B_{i} $ and $\bbs_{j}^{G_{2}} \subseteq B_j$  are quadrature points chosen by this method then the covariance entry $\bbK_{i,j}$ is approximated  by  
More precisely, given grids $G_{i}$ and $G_{j}$ tailored to regions $B_{i}$ and $B_{j}$ respectively, with $\bbs_i^{G_i} = \{\bbs: \bbs\in B_{i}\cap G_i\},\ i=1,\ldots,n$, with $L_i^{G_i}$ the number of points in $\bbs_i^{G_i}$, then the covariance entry $\bbK_{i,j}$ is approximated  by
 \begin{eqnarray}
 %\frac{1}{|B_i|} \frac{1}{|B_j|} \sum_ {k,l} c_Y(\bbu_k^m, \bbv_l^n) \boldsymbol{\Delta}
 %\frac{1}{\cC(\bbs_i^{G_{1}})} \frac{1}{\cC(\bbs_j^{G_{2}})} \sum_ {\bbs_{k}\in \bbs_i^{G_{1}},\bbs_{\ell}\in \bbs_j^{G_{2}}}  c_Y(\bbs_k, \bbs_\ell) \boldsymbol{\Delta}.
\frac{1}{L_i^{G_i} L_j^{G_j}} \sum_ {\bbs_{k}\in \bbs_i^{G_{i}},\bbs_{\ell}\in \bbs_j^{G_{j}}}  c_Y(\bbs_k, \bbs_\ell).
  \label{eqJR}
 \end{eqnarray}

\noindent The recommended maximum grid densities to use for the grids $G_{i},  i=1,\ldots,n$ with the [JH] approach are 10 points for a one-dimensional domain, $6 \times 6$ for a two-dimensional domain, and $4 \times 4 \times 4$ for a three-dimensional domain. This approach is particularly well-suited to parallelogram regions, as the quadrature point locations can be centered in each unit of a regular partitioning of the overall region, avoiding the potential for bias that arises when more geometry-agnostic regular grid overlay methods are applied.\\

\vskip-1em
\noindent The [JH] approach, however, has several disadvantages for large data sets.  The number of points in this method is limited to avoid generating large sums over covariance matrices in Eqn.~(\ref{eqJR} ).  This will be an issue for large numbers of locations because in practice evaluating transcendental functions for the covariance kernel will take appreciable computation time. In restricting the number of quadrature points, however,  this  results in limited accuracy for irregular shapes. To improve representing the integral the [JH] method also proposes quadrature points aligned with the extent and boundaries of the regions. This makes coding this method complicated.  Finally we note that because the set of quadrature points  is adapted to each region's shape and locations, it makes it difficult to describe an underlying discrete process that will be the basis for prediction. \\
 
\vskip-1em
\noindent Our approach  revisits  the Riemann double sum over a large fixed grid. However, by taking advantage of the fast Fourier transform and stationarity of $c_Y$, we are able to reduce the double sum over the grid to a single sum in frequency space. In addition, there is the possibility of even more rapid evaluation of the covariance function if it has a closed form Fourier transform, such as the Mat\`{e}rn family. 

\subsection{FAIR algorithm }\label{subsec:FAIRalgo}

We seek to evaluate integrals of the type shown in Eqn.~(\ref{eq:target}) using Fourier representations. Given this strategy it is 
useful to review some of this area. Assume a stationary 
covariance model $c_Y(\bbs_i,\bbs_j) = c(\bbs_i - \bbs_j)$. Let $\ast$ denote the (multi-dimensional) convolution operator, 
overlines denote complex conjugates, and $\cF \left[g \right] (\bomega)$ denote the (multi-dimensional) Fourier transform 
of the function $g(\bbs)$.  We first cast the computation of $\bbK_{i,j}$ in terms of this continuous transform and then 
approximate this representation using the DFT. \\

\vskip-1em
\noindent We make use of the convolution theorem,
\begin{equation}
    f = g \ast h \Leftrightarrow \cF \left[f \right] =\cF \left[g \right]   \cF \left[h \right].  \nonumber
\end{equation}

\noindent Also, by Plancherel's Theorem, 
\begin{equation}
    \int_{\bbR^d} f(\bbs)  \overline{g(\bbs)} d\bbs = 
    \int_{\bbR^d} \cF \left[f \right] (\bomega)  \overline{\cF \left[g \right] (\bomega)} d\bomega. \nonumber
\end{equation}

\vskip-1em
\noindent With these results we have,
\begin{equation}
\begin{aligned}
\int_{B_i} \int_{B_j} c_Y(\bbu,\bbv) d\bbu d\bbv  & = 
 \int_{\bbR^d} I_{B_i}(\bbu)   ( I_{B_j} \ast c ) (\bbu) d\bbu
 \\ & =  \int_{\bbR^d} \cF \left[I_{B_i} \right]  (\bomega)  
 \overline{ \cF \left[ I_{B_j} \ast c  \right] (\bomega) } d\bomega
  \\ & =  \int_{\bbR^d} \cF \left[ I_{B_i} \right] (\bomega) 
    \overline{ \cF \left[ I_{B_j} \right] (\bomega)  }
  \cF \left[ c \right] (\bomega)  d\bomega, 
  \label{eq:planch}
  \end{aligned}
\end{equation}
\noindent where, $I_{A}(\bbx)=1\ \mbox{if}\ \bbx\in A\ \mbox{and}\ 0,\ \mbox{otherwise}$. Furthermore, the representation given in Eqn.~(\ref{eq:planch}) is exact. \\

\begin{figure}[!h]
\centering
\includegraphics[width = 0.5\textwidth]{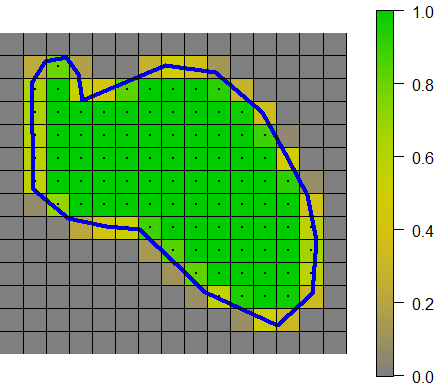}
\caption{Example output of $\mathcal{I}$ for an arbitrary region and grid.}
\label{fig:frac_ind}
\end{figure}

\vskip-1em
\noindent We now make the following two approximations to the final expression in Eqn.~(\ref{eq:planch}). 
\begin{itemize}
\item[(a)] The first approximation is the restriction of the integral to a finite domain. Note that, for functions $c(\bbs)$ such that $c(\bbs)\rightarrow 0$ as $||\bbs||\rightarrow \infty$, $( I_{B_i} \ast c )$ will also decay to zero away from $B_i$; thus, we select $S \subset \bbR^d$, a $d$-dimensional, rectangular region (e.g., an interval in $\bbR^1$, a rectangle in $\bbR^2$, a rectangular prism in $\bbR^3$, etc.), where $B_i,B_j$ are well contained in $S$ (i.e., the distance between $B_i,B_j$ and the boundaries of $S$ is large, relative to the decay range of $c(\cdot)$).  Accordingly this approximation will restrict the integral in Eqn.~(\ref{eq:planch}) to the domain $S$. 

\item[(b)] The second approximation is a discretization. 
Let $G$ be a regular grid of points that covers $S$ and with spacing $\Delta_j$ in dimension $j$. For a function, say $g$, evaluated on this grid, let $DFT[ g]$ denote its DFT. Let $\omega^G$ be the mirror grid in the frequency domain.  Then we are led to the approximation:

\begin{eqnarray}
\label{eq:disc_approx}
% \bbK_{i,j}  \approx  \frac{1}{\cC(\bbs_i^{G})\cC(\bbs_j^{G})}\sum_k   DFT[{\mathcal{I}_{B_{i}}}](\omega_k^G)     \overline{    DFT[{ \mathcal{I}_{B_{j}}}](\omega_k^G)  } DFT[{c}](\omega_k^G)  \Delta,
\int_{B_i} \int_{B_j} c_Y(\bbu,\bbv) d\bbu d\bbv \approx  \sum_k   DFT[{\mathcal{I}_{B_{i}}}](\omega_k^G)     \overline{    DFT[{ \mathcal{I}_{B_{j}}}](\omega_k^G)  } DFT[{c}](\omega_k^G)  \Delta
\end{eqnarray}
where $\Delta = \Pi_{i=1}^{d}\Delta_i$, and $\mathcal{I}_{B_{i}}$ is a weighted version of the indicator function $I_{B_i}$. Each grid point can be associated with a surrounding grid box, and $\mathcal{I}_{B}$ at a grid point is equal to the fraction of the area of the associated grid box that is contained in $B$. Values for $\mathcal{I}_{B}$ will be mostly zero or one with fraction values for grid boxes on the boundary of $B$. In two dimensions one can evaluate an approximate version of $\mathcal{I}_{B}$ rapidly, and an example is illustrated in Figure~\ref{fig:frac_ind}.
\end{itemize}

\subsection*{Implementation}
Our method approximates the integrals shown in Eqn.~(\ref{eq:planch}) with discretized sums over a finite domain as shown in Eqn.~(\ref{eq:disc_approx}), exploiting the fast Fourier transform (FFT) algorithm to compute the DFT. The algorithm to determine an entry of $\bbK$ is as follows. Throughout this algorithm we will use $\theta_x$ to refer to the distance where the correlation function decreases to $x$. %For a given $x$, this is typically a constant multiple of the scale parameter (or range) parameter in a covariance function. 

\begin{enumerate}
    \item A regular grid $G$ is constructed which extends beyond the regions ${B_1,...,B_n}$. The overall extent of the grid is constrained to be at least twice $\theta_{0.05}$ and also extend beyond all observation regions by at least $\theta_{0.25}$.
 The grid size is chosen to be highly composite  (dyadic) to facilitate an efficient FFT.  
    \item $\widehat{\bbB} ^i = DFT[\mathcal{I}_{B_i }] $ is found by FFT for $i=1,\dots,n$.    
    \item $\widehat{\bbc} = DFT[c]$ is found for $c(\cdot)$ by FFT. 
%    \item $\bbK_{i,j} =  \frac{1}{\cC(\bbs_i^{G})\cC(\bbs_j^{G})}  \sum_k \widehat{\bbB}^i_k  \overline{\widehat{\bbB}^j_k} \widehat{\bbc}_k  \Delta $ are found with the sum being over the discretized Fourier frequencies prescribed by the DFT. 
   \item $\widehat{|B_i|}=\sum \mathcal{I}_{B_{i}}(G) \Delta$ is found for $i=1,\dots,n$.
   \item $\bbK_{i,j} =  \frac{1}{\widehat{|B_{i} |}} \frac{1}{\widehat{|B_{j} |}}  \sum_k \widehat{\bbB}^i_k  \overline{\widehat{\bbB}^j_k} \widehat{\bbc}_k  \Delta $ are found with the sum being over the discretized Fourier frequencies prescribed by the DFT. 
\end{enumerate}

\noindent The grid extent constraints are motivated by two factors. The discrete, finite domain approximation to the full Fourier transform of $c$ degrades significantly when the grid is constructed over a domain which is small relative to the decay rate of the covariance function. This is due to the well known property of the DFT as enforcing a periodic representation of the covariance function. %For example, experiments show a sharp reduction in quality if a spatial domain dimension is less than twice the 0.10 correlation range of $c$. 
Extending the grid to be at least twice $\theta_{0.05}$ avoids this issue. Also, the discrete, finite domain approximation degrades significantly without  a `buffer zone' around the regions ${B_1,...,B_n}$ based on correlation range. This is due to the periodic wrapping of the approximation to the covariance whereby regions on one edge of the domain become (spuriously) correlated with regions on the opposite edge. Experiments show that extending the grid with a buffer of $\theta_{0.25}$ avoids this artifact. Thus, if the spatial domain that contains the regions ${B_1,...,B_n}$ is a square with a length $L$, a conservative grid for computation should be extended to have extent of at least $ max( L + 2 \theta_{0.25}, 2 \theta_{0.05}) $. The larger the extent of the grid, the more these effects are minimized, but larger extents at a fixed grid resolution also result in coarser evaluations of the functions in question.\\

\vskip-1em
\noindent The accuracy of this method is also a function of the resolution of the grid. In Section~\ref{subsubsec:acc_method} we provide a numerical study to suggest the relationship between grid spacing ($\Delta$), extent, and accuracy.

\subsection{Surface Prediction} \label{subsec:surf_pred}
\noindent The proposed FAIR algorithm also provides intermediate quantities to make surface prediction (see, Eqn.~(\ref{eq:target00})) efficient. In particular, the fine grid $G$ (containing $L$ points), used for the computation of the change of support integrals, can be reused for creating a predicted surface.  To simplify the exposition we will assume throughout this section  that $\bbmu, \mu_{Y}^{\cdot}, \tau^{2}$ and any additional parameters in $c_{Y}$ are known. Now, motivated by Eqn.~(\ref{eq:target00} ) one can define the vector
 \[ \bbeta=  ( \bbK + \tau^2 \bbI ) ^{-1}(\bbZ - \bmu).  \]
Now focusing on the second term of Eqn.~(\ref{eq:target1}) and an arbitrary $i$-th member of the grid, denoted by $\bbu_{i}$, we have:
\begin{eqnarray*}
\bbk(\bbu_{i})^T\bbeta &=&  \sum_{\ell=1}^n  \left( \frac{1}{|B_{\ell}|}\int_{B_\ell} c_Y(\bbu_{i},\bbu) d\bbu \right)  \bbeta_{\ell} 
= \int   \sum_{\ell=1}^n \left( \frac{1}{|B_{\ell}|} I _{B_\ell}(\bbu) \bbeta_\ell \right)  c_Y(\bbu_{i},\bbu) d\bbu\\
&=& \int   \phi(\bbu)  c( \bbu_{i} - \bbu) d\bbu =  (\phi \ast c)[ \bbu_{i}],
\end{eqnarray*}
where  $\phi( \bbu) = \sum_{\ell=1}^n  \frac{ \bbeta_\ell}{|B_{\ell}|} I _{B_\ell} (\bbu) $. 
Note that the last expression is a convolution of $\phi$ with $c$ evaluated at $\bbu_i$; using the same discrete approximation described above we are lead to 
$$ \widehat{\bbY}^G = \mu_{Y}(G) + DFT^{-1} [ DFT[\phi] DFT[ c] ], $$
where $DFT(\phi)$ and $DFT(c)$ have support on the mirror grid in the frequency domain, and $\mu_{Y}(G)=\{\mu_{Y}(\bbu_i) \}_{i=1}^{L}$.  Note that the $i^{th}$ component of $\widehat{\bbY}^G$ is the predicted value at $\bbu_i$.  Thus, evaluation of the predicted surface on the entire grid $G$ is $\widehat{\bbY}^G$, obtained in a single step using the inverse DFT. Additionally, note that $DFT[\phi] = \sum_{\ell=1}^n \frac{ \bbeta_\ell}{|B_{\ell}|} DFT[I _{B_\ell}] \approx \sum_{\ell=1}^n \frac{ \bbeta_\ell}{\widehat{|B_{\ell}|}} DFT[\mathcal{I}_{B_\ell }] $, and from evaluation of $\bbeta$, $\widehat{|B_\ell|}$ as well as $DFT[\mathcal{I}_{B_\ell }] $ for each $B_\ell$ and $DFT[c]$ have already been computed. \\

\section{Numerical Studies}\label{sec:numstudy}

In this section, we consider a zero-mean spatial process $Y(\bbs)$ with covariance function $c_{Y}(\cdot)$. In general, integrals of the form in Eqn.~(\ref{eq:target}) are not subject to analytical solution over irregular regions $B_i,B_j$ and common covariance functions $c_Y$, but there are exceptions. In Section~\ref{subsubsec:acc_method}, we consider a special case of conveniently selected regions and Gaussian covariance function $c_Y(\mathbf{h})=\exp{[-\|\mathbf{h}\|^2/2]}$ that allow a ``ground truth'' that can be obtained to double precision floating point accuracy. We then use this ground truth to determine the accuracy of FAIR relative to changes in grid resolution and distance between regions. In Section~\ref{subsec:mat_method}, we consider a Mat\'ern covariance function (with marginal variance $\sigma^{2} = 1$, range $\theta = 0.5$, and smoothness $\nu = 1.5$) to show that for a commonly used covariance model over a modest number of irregular regions of interest, the estimates generated by FAIR are of equal quality to those generated by a simple Riemann sum approach (for a given grid resolution),  but as the grid resolution increases, the computational expense of FAIR scales more favorably than the expense of the Riemann sum approach. Finally, in Section~\ref{subsec:census}, we perform an example estimation and prediction process on census block data using both FAIR and Riemann approaches to provide a comparison in the context of an application.

\subsection{Gaussian Covariance Accuracy Study}\label{subsubsec:acc_method}
In this section, we consider the underlying process $Y(\bbs)$ being averaged over two rectangular regions with sides parallel to the coordinate axes. In particular, the regions are $A=[ax_{1},ax_{2}]\times [ay_{1},ay_{2}]$ and $B=[bx_{1},bx_{2}]\times [by_{1},by_{2}]$, respectively. Then straightforward calculation yields,
\begin{multline}
\int_A \int_B c_Y(\mathbf{u},\mathbf{v}) d\mathbf{v} d\mathbf{u}= \\ \left ( - \int_{bx_2-ax_1}^{bx_2-ax_2} \Phi(w_1) dw_1 + \int_{bx_1-ax_1}^{bx_1-ax_2} \Phi(w_2) dw_2 \right ) \left ( - \int_{by_2-ay_1}^{by_2-ay_2} \Phi(w_3) dw_3 + \int_{by_1-ay_1}^{by_1-ay_2} \Phi(w_4) dw_4 \right ), \label{eq:intermed_accuracy}
\end{multline}
where $\Phi(.)$ denotes the cumulative distribution function of a Normal(0,1) random variable. Note that the integrals in Eqn.~(\ref{eq:intermed_accuracy}) are of the same general type and can be computed using integration by parts.\\

\vskip-1em
\noindent For our accuracy studies, we consider pairs of unit squares $A = [0,1]^{2}$ and $B_{i} = [\delta_{i},1+\delta_{i}]^{2}$, where the offset $\delta_{i}$ is a separation distance between the regions.
For each $\delta_{i}$, the ground truth covariance matrix for the two averaged regional values was computed. The ground truth correlation between regions varies with their separation distance - some representative values are shown in Table \ref{tab:acc_study_true_corr}. \\

\begin{table}[!h]
\begin{center}
\begin{tabular}{|c|ccccc|}
\hline
$\delta_{i}$      & 0.3   & 0.9   & 1.5   & 2.1   & 2.7   \\ \hline
$Cor(z_{A},z_{B_i})$ & 0.926 & 0.501 & 0.147 & 0.023 & 0.002 \\ \hline
\end{tabular}
\caption{Gaussian accuracy studies: ground truth correlations by separation distance.}
\label{tab:acc_study_true_corr}
\end{center}
\end{table}

\begin{figure}[!h]
\centering
\includegraphics[width = 0.8\textwidth]{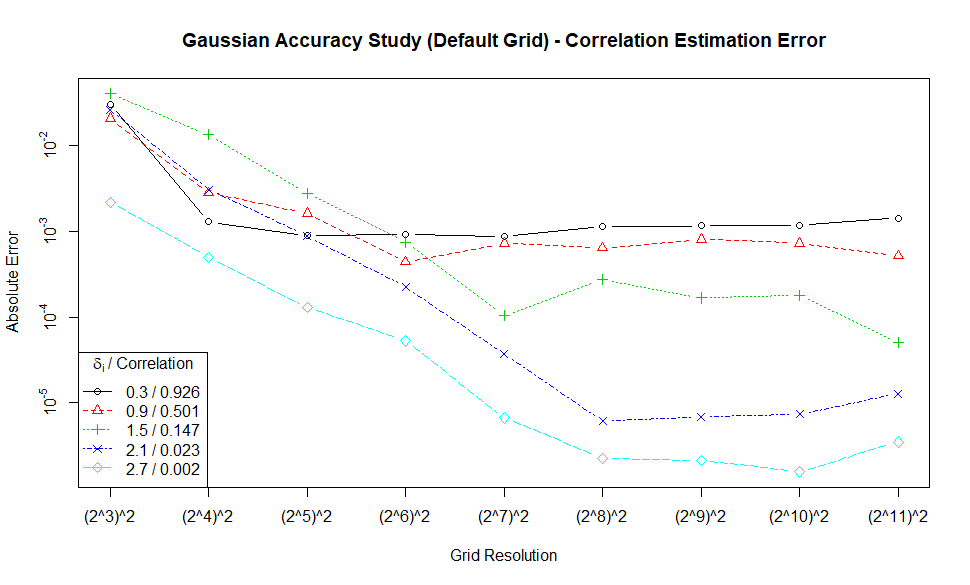}
\caption{Accuracy for Gaussian covariance function with default grid extent.}
\label{fig:acc_results_default}
\end{figure}

\vskip-1em
\noindent For each, we employ FAIR to approximate the covariance matrix using varying grids (in each case, using the default image radius and region padding selection method as described in Section~\ref{subsec:FAIRalgo}. The grid resolution was varied over integer powers of two: the coarsest grid consisted of $2^3 \times 2^3 = 8 \times 8 = 64$ total grid points, while the finest grid had $2^{11} \times 2^{11} = 2048 \times 2048 \approx$ 4 million points. The approximated covariance matrices are used to generate correlations, which are then compared against the ground truth values. The absolute errors observed during these studies are shown in Figure~\ref{fig:acc_results_default}.\\

\begin{figure}[!h]
\centering
\includegraphics[width = 0.8\textwidth]{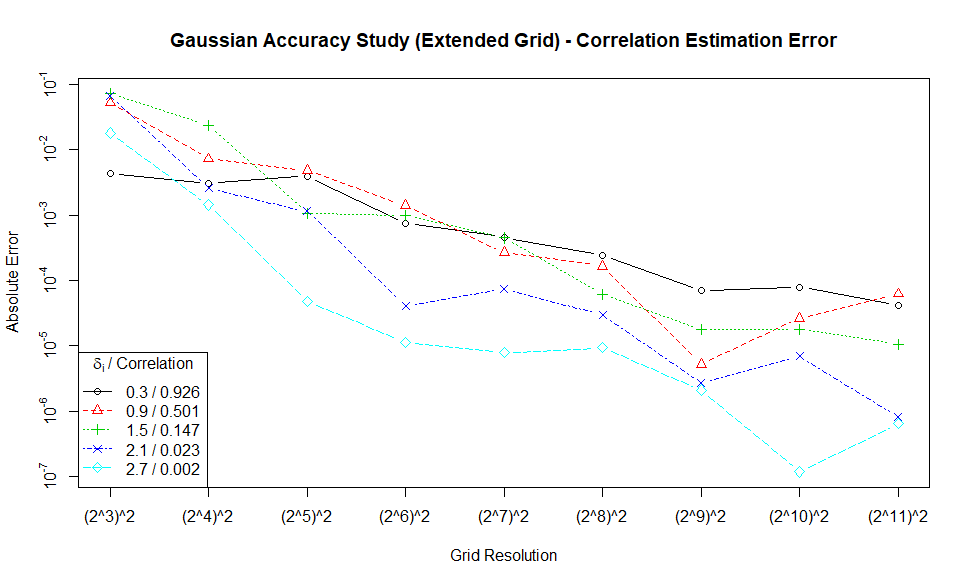}
\caption{Accuracy for Gaussian covariance function with expanded grid extent.}
\label{fig:acc_results_extended}
\end{figure}

\vskip-1em
\noindent Overall, our method does an excellent job in terms of accuracy and as one would expect, accuracy improves with increased grid resolution, until hitting a plateau. Note that shape pairs with smaller offsets reach this plateau at lower resolutions than pairs with larger offsets. This plateau represents the transition from accuracy being constrained by grid resolution to being constrained by grid extent. When the study is repeated with larger grid extents (in this instance, increasing default extent constraints by a single unit, i.e., extent set to $ max( L + 2 \theta_{0.25}+1, 2 \theta_{0.05}+1)$), the plateau appears at lower error values/higher grid resolutions, as shown in Figure~\ref{fig:acc_results_extended}, but there is also a decrease in accuracy at lower resolutions for most of the offset distances. This loss of accuracy is due to coarser grids (spreading the same number of grid points over a larger extent), which then lead to coarser evaluations of the functions involved.

\subsection{Mat\'ern Covariance Timing and Consistency Study} \label{subsec:mat_method}
\subsubsection{Study Design} \label{subsubsec:tim_method}
\begin{figure}[!h]
\centering
\includegraphics[width = 0.5\textwidth]{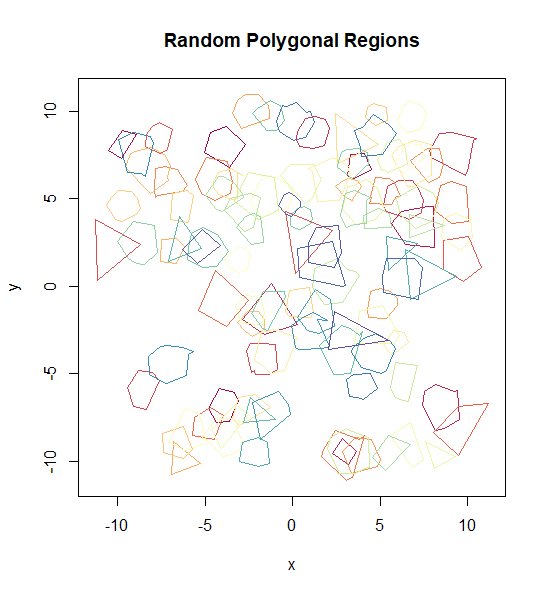}
\caption{Randomly-generated irregular polygonal regions of interest employed in Mat\'ern covariance timing study.}
\label{fig:tim_study_regions}
\end{figure}

In this section, we demonstrate the degree to which the computational cost of our method scales more favorably than direct quadrature, and in a more realistic context. We consider regions that have a random polygon shape and the Mat\'ern family of covariance functions. We fix a square spatial domain with dimensions $[-11,11]^2$. We generate one hundred random polygons with approximate centers drawn from a uniform distribution on $[-10,10]^2$.  The number of sides, sizes and irregularity of the sides are all randomly chosen with self-intersecting polygons being omitted. The realized polygons are shown in Figure~\ref{fig:tim_study_regions}.\\

\vskip-1em
\noindent  For an underlying process with Mat\'ern covariance (with range 0.5, smoothness 1.5, and marginal variance 1.0), direct quadrature and FAIR were used to estimate the $100\times100$ covariance matrix associated with these regions using grids with different resolutions covering the domain $[-12.5, 12.5]^{2}$. We consider grid resolutions $2^{m},m=7,8,9,10$. For each resolution considered, measures of the consistency between the estimates were collected, as well as timing results for each case.\\
 
\vskip-1em

\noindent Note that in the direct quadrature case, each entry in the target matrix computed as in Eqn.~\ref{eqRieman} effectively requires the evaluation of the covariance function over a matrix of distances that is $m \times n$, where $m,n$ are the number of grid points for which the two regional indicator functions $ I_{B_k}( \bbs_k^G)$ and $I_{B_\ell}( \bbs_{\ell}^G)$ are indicating. For very large grids, these matrices may exceed what can be stored in memory, and require additional costs to compute piecemeal. However, for the resolutions considered in this study, the largest of these matrices was approximately $8000\times 8000$ and thus was not affected by storage issues.\\
 
\vskip-1em

\noindent Additionally, it is worth noting that because the regions of interest are polygons with a small number of sides, one may obtain more accurate integrations by a triangularization of the regions and numerical quadrature. The goal of this study is not to suggest that either method examined here is the very best possible for this particular problem; rather to compare the two against each other in a setting more complex than simple regular rectangles. Both algorithms tested here (namely, FAIR and direct quadrature) do not require triangularization of regions - both require that the regions of interest be reasonably representable by indicator functions on a regular grid.\\
 
\vskip-1em

\noindent We perform the entire timing study using an i7-7700 @ 3.60GHz (w. 16GB RAM). Because there is not an analytic ground truth benchmark available in this case, absolute accuracy measures are not available, so several comparative measures are provided.

\subsubsection{Results}
\label{subsubsec:time_results}
\par We consider the following three criteria to assess the agreement between approximated covariance matrices $\widehat{\Sigma}_{FAIR}$ and $\widehat{\Sigma}_{Direct}$, generated by the two methods, FAIR and Direct, respectively, at each resolution.
\begin{itemize}
\item[(a)] \emph{The root mean squared entry-wise difference (RMSED):}~ \\ $RMSED(\widehat{\Sigma}_{FAIR}, \widehat{\Sigma}_{Direct}) = \frac{1}{n}\sqrt{\sum_{i=1}^{n}\sum_{j=1}^{n} (\widehat{\Sigma}_{FAIR;~i,j}-\widehat{\Sigma}_{Direct;~i,j})^2}$.
\item[(b)] \emph{The maximum absolute entry-wise difference (MAED):}~ \\ $MAED(\widehat{\Sigma}_{FAIR}, \widehat{\Sigma}_{Direct}) = \max_{i,j \in \{1\ldots n\}} \| \widehat{\Sigma}_{FAIR;~i,j}-\widehat{\Sigma}_{FAIR;~i,j} \|$.
\item[(c)] \emph{The Kullback-Leibler (KL) divergence:}~ \\ $D_{KL}(\widehat{\Sigma}_{FAIR}, \widehat{\Sigma}_{Direct})=\frac{1}{2}\left(\mbox{Tr}[\widehat{\Sigma}_{FAIR}^{-1}\widehat{\Sigma}_{Direct}]-n+\log{\left(\frac{\det{\widehat{\Sigma}_{FAIR}}}{\det{\widehat{\Sigma}_{Direct}}}\right)}\right).$
\end{itemize}
\noindent In all of the above, $n$ is the number of regions ($n=100$ in this study). The $RMSED$ and $MAED$ measures are chosen to provide a sense of how the two estimates differ on an entry-by-entry basis, while $D_{KL}$ provides a sense of how they may differ in the context of a likelihood evaluation. For all three measures, smaller values indicate more similar matrices.\\
 
\vskip-1em

\noindent Timing results and difference metrics at each resolution considered are shown in Table~\ref{tab:tim_study_tim_sim}. The results show that the approximations generated by the two methods become more similar as resolutions increase. The results also demonstrate the efficient scaling of the FAIR method; note that as the grid resolution doubles, execution times for the FAIR method increase by a factor of approximately 2.5; contrast that with a factor of approximately 16 for the direct method.\\

\begin{table}[!h]
\begin{center}
\begin{tabular}{|c|c|c|c|c|c|c|}
\hline
Resolution    & $\Delta x$ & $RMSED(\widehat{\Sigma}_{F}, \widehat{\Sigma}_{D})$   & $MAED(\widehat{\Sigma}_{F}, \widehat{\Sigma}_{D})$    & $D_{KL}(\widehat{\Sigma}_{F}, \widehat{\Sigma}_{D})$  & Time: FAIR       & Time: Direct   \\ \hline
$2^7 \times 2^7$   & 0.194                     & 8.168e-03 & 8.511e-02 & 1.701e+01 & 11.4s           & \textbf{8.2s} \\ \hline
$2^8 \times 2^8$   & 0.097                     & 4.023e-03 & 4.284e-02 & 3.078e+00 & \textbf{27s}  & 128s        \\ \hline
$2^9 \times 2^9$   & 0.048                     & 2.026e-03 & 2.169e-02 & 7.107e-01 & \textbf{66s}  & 2099s       \\ \hline
$2^{10} \times 2^{10}$ & 0.024                     & 1.039e-03 & 1.124e-02 & 1.792e-01 & \textbf{191s} & 33755s      \\ \hline
\end{tabular}
\caption{Observed timing and approximation similarity measures in the Mat\'ern covariance timing study, comparing $\widehat{\Sigma}_{FAIR}$ to $\widehat{\Sigma}_{Direct}$ at each grid resolution (denoted $\widehat{\Sigma}_{F}$ and $\widehat{\Sigma}_{D}$ respectively).}
\label{tab:tim_study_tim_sim}
\end{center}
\end{table}

\vskip-1em
\noindent The estimates from each method at each resolution are also compared to the direct method at the highest resolution. These results are shown in Tables ~\ref{tab:tim_study_hr_RMSED}, \ref{tab:tim_study_hr_MAED}, and \ref{tab:tim_study_hr_DKL}.\\

\begin{table}[!h]
\begin{center}
\begin{tabular}{|c|c|c|}
\hline
Grid Resolution        & $RMSED(\widehat{\Sigma}_{F}, \widehat{\Sigma}_{HR})$ & $RMSED(\widehat{\Sigma}_{D}, \widehat{\Sigma}_{HR})$ \\ \hline
$2^7 \times 2^7$	& 8.170e-03	& 1.158e-03 \\ \hline
$2^8 \times 2^8$	& 4.032e-03	& 3.368e-04 \\ \hline
$2^9 \times 2^9$	& 2.027e-03	& 1.331e-04 \\ \hline
$2^{10} \times 2^{10}$	& 1.039e-03	& - \\ \hline
\end{tabular}
\caption{Observed $RMSED$ values in the Mat\'ern covariance timing study, comparing $\widehat{\Sigma}_{FAIR}$ and $\widehat{\Sigma}_{Direct}$  at each grid resolution to $\widehat{\Sigma}_{Direct}$ at the highest grid resolution (denoted $\widehat{\Sigma}_{HR}$).}
\label{tab:tim_study_hr_RMSED}
\end{center}
\end{table} 

\begin{table}[!h]
\begin{center}
\begin{tabular}{|c|c|c|}
\hline
Grid Resolution        & $MAED(\widehat{\Sigma}_{F}, \widehat{\Sigma}_{HR})$ & $MAED(\widehat{\Sigma}_{D}, \widehat{\Sigma}_{HR})$ \\ \hline
$2^7 \times 2^7$	& 8.494e-02	& 1.978e-02 \\ \hline
$2^8 \times 2^8$	& 4.321e-02	& 6.720e-03 \\ \hline
$2^9 \times 2^9$	& 2.160e-02	& 2.487e-03 \\ \hline
$2^{10} \times 2^{10}$	& 1.124e-02	& - \\ \hline
\end{tabular}
\caption{Observed $MAED$ values in the Mat\'ern covariance timing study.}
\label{tab:tim_study_hr_MAED}
\end{center}
\end{table}

\begin{table}[!h]
\begin{center}
\begin{tabular}{|c|c|c|}
\hline
Grid Resolution        & $D_{KL}(\widehat{\Sigma}_{F}, \widehat{\Sigma}_{HR})$ & $D_{KL}(\widehat{\Sigma}_{D}, \widehat{\Sigma}_{HR})$ \\ \hline
$2^7 \times 2^7$	& 1.675e+01	& 1.488e-01 \\ \hline
$2^8 \times 2^8$	& 3.091e+00	& 2.119e-02 \\ \hline
$2^9 \times 2^9$	& 7.145e-01	& 1.766e-03 \\ \hline
$2^{10} \times 2^{10}$	& 1.792e-01 & - \\ \hline
\end{tabular}
\caption{Observed $D_{KL}$ values in the Mat\'ern covariance timing study.}
\label{tab:tim_study_hr_DKL}
\end{center}
\end{table}

\vskip-1em
\noindent In this experiment, more than half of the total execution time for the FAIR algorithm was taken up by the formation of the weighted indicator functions. Our implementation of these functions is presently at proof-of-concept stage, and we expect that significant timing improvements are still possible in this portion of the algorithm. In some contexts (e.g. numerical maximum likelihood estimation), this cost (as well as the calculation of the Fourier transforms of these functions) may be considered a one-time expense, as varying only the covariance kernel (using a fixed set of regions and a fixed grid) does not require recomputation of these functions, and in Section \ref{subsec:census}, these costs are handled in just such a fashion. Additionally, if very high resolution grids are used, then the accuracy benefits of using an improved indicator function on such grids may be marginal, and the more typical "in or out" functions may be used at some savings.

\subsection{Dataset Example} \label{subsec:census}

\noindent  In this section, we demonstrate the performance of the algorithm with a spatial dataset, specifically a spatial estimation and prediction problem using real-world data and region geometry.\\
 
\vskip-1em

\begin{figure}[!h]
\centering
\includegraphics[width = 0.8\textwidth]{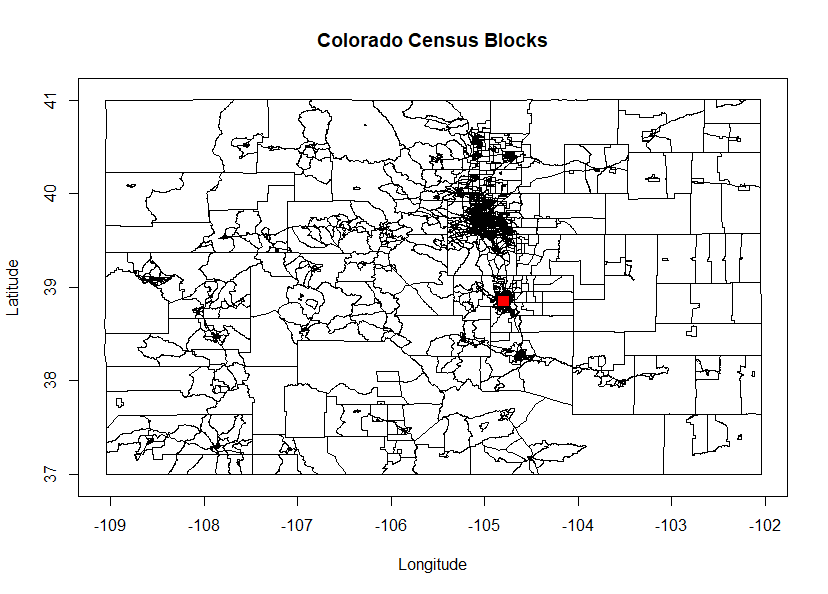}
\caption{Census block geometry for Colorado state, with dataset example subset indicated.}
\label{fig:co_blocks}
\end{figure}

\noindent The region geometry selected for this study are the 84 US Census blocks entirely contained in the region that extends from 104.85 to 104.75 West Longitude, and from 38.8 to 38.9 North Latitude (a region covering some of Colorado Springs, CO) as indicated in Figure~\ref{fig:co_blocks}. The data chosen are the block-level counts of traffic intersections (an indicator of economic development). Intersection data made available by Patricia Romero Lankao from the National Energy Renewable Laboratory and aligned with the 2016 census block group shape files, obtained from  \url{https://www2.census.gov/geo/tiger/} as  GENZ2016 2016 Cartographic Boundary Files  (cb\_2016\_08\_bg\_500k)\footnote{This data set is compiled by The Center for Neighborhood Technology (\url{cnt.org}) and distributed in support of the Housing and Transportation (H+T) Affordability Index. State by state  files in \texttt{cvs} format are available for download from \url{https://htaindex.cnt.org/}}.\\

\vskip-1em

\begin{figure}[!h]
\centering
\includegraphics[width = 0.6\textwidth]{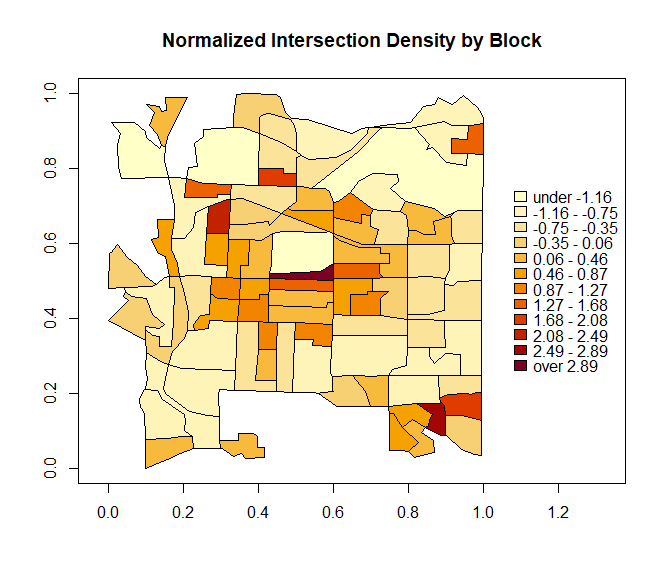}
\caption{Rescaled, normalized block geometry and intersection density data employed in dataset example.}
\label{fig:cs_norm_data}
\end{figure}

\noindent We shift and rescale the region geometry to occupy a unit square at $[0,1]^2$, transform the count data to densities by dividing by block area, and then normalize the density data using the observed sample mean and standard deviation. The transformed data are shown in Figure~\ref{fig:cs_norm_data}.\\

\vskip-1em

\noindent We model the normalized data as block averages of an underlying Gaussian process with a Mat\'ern covariance. We assume a smoothness of 1.5, and estimate the marginal variance, range, and (post-aggregation) nugget variance using maximum likelihood methods - to facilitate the timing comparison, we concentrate the marginal variance and nugget variance into a ratio term, and perform a simple grid search over (range parameter, variance ratio) candidate pairs - 9 distinct range candidates were considered between 0.0625 and 0.50, with 45 distinct variance ratio candidates between 0.005 and 2.0. FAIR was implemented with a $2^{10} \times 2^{10}$ grid extending approximately to $[-2,3]^2$. Riemann quadrature points were subsampled from the FAIR grid, ranging from163 points in the smallest region, up to 2593 points in the largest region, with a mean of 414 points per region.\\
 
\vskip-1em

\noindent 
Each range parameter candidate investigated requires the recalculation of the regional covariance matrix. For each candidate, these calculations were repeated using both the FAIR algorithm and the standard Riemann one. Because the region geometry does not vary across these calculations, some quantities can be precomputed (the Fourier transforms of the fractional indicator functions in the case of FAIR, and the coordinates of the quadrature points in each region in the case of Riemann). For larger range parameter candidates, some matrices returned by FAIR were not quite positive definite, and were thus coerced to the nearest positive definite matrix using \texttt{Matrix::nearPD} in \texttt{R}. Additionally, the geometry of these regions (while still polygonal) is more complex than in our previous studies, with a mean side count of 12. These factors contribute to differing cost ratios in this study, relative to those observed in the study described in Section~\ref{subsec:mat_method}. Timing results for the major phases of each algorithm for this estimation task are provided in Tables \ref{tab:FAIR_pred} and \ref{tab:Riemann_pred}. \\

\begin{table}[!h]
\begin{center}
\begin{tabular}{|l|l|}
\hline
\multicolumn{2}{|c|}{\textbf{Performance - FAIR Algorithm}} \\ \hline
Setup Grid                             & 0.14s      \\ \hline
Precompute Indicator Functions and Fourier Transforms                & 33.39s     \\ \hline
Maximum Likelihood Estimation             & 1022.44s   \\ \hline
\textbf{Total Execution Time}                  & 1055.97s   \\ \hline
\end{tabular}
\caption{Timing results observed for FAIR algorithm during estimation task for dataset example.}
\label{tab:FAIR_pred}
\end{center}
\end{table}

\begin{table}[!h]
\begin{center}
\begin{tabular}{|l|l|}
\hline
\multicolumn{2}{|c|}{\textbf{Performance - Riemann Algorithm}} \\ \hline
Setup Quadrature Points                   & 3.73s       \\ \hline
Maximum Likelihood Estimation               & 2516.25s    \\ \hline
\textbf{Total Execution Time}                    & 2519.98s    \\ \hline
\end{tabular}
\caption{Timing results observed for Riemann algorithm during estimation task for dataset example.}
\label{tab:Riemann_pred}
\end{center}
\end{table}

\vskip-1em
\noindent With covariance parameter estimates found, we use the values computed by the FAIR algorithm during the estimation process to predict the underlying surface on the default grid. As noted in Section~\ref{subsec:surf_pred}, this involves only a weighted sum of the Fourier-transformed indicator functions to form $DFT[\phi]$, an entry-wise product of the $DFT[\phi]$ and $DFT[c]$ terms, and an inverse FFT, and incurs only 2.02s of additional compute time. A subset of the predicted surface is shown in Figure~\ref{fig:cs_norm_data} (only a subset is provided, as away from the observation regions, the prediction reverts to the mean). \\

\begin{figure}[!h]
\centering
\includegraphics[width = 0.6\textwidth]{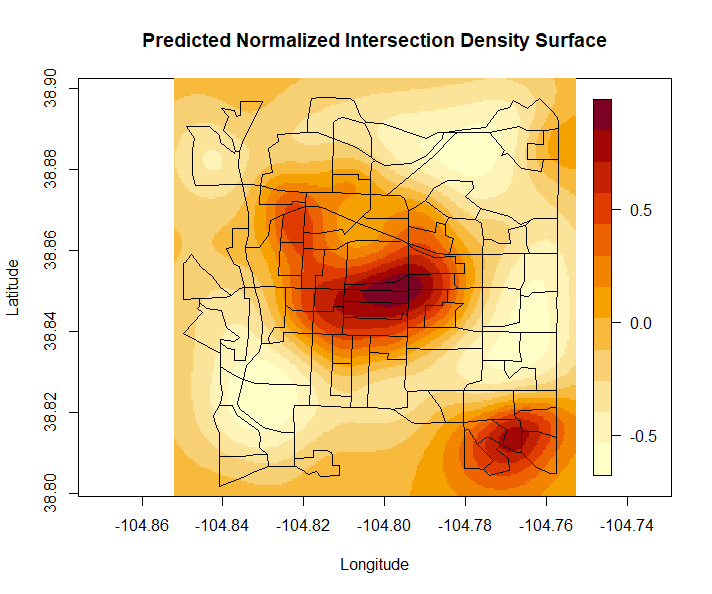}
\caption{Surface prediction generated using FAIR algorithm products in dataset example.}
\label{fig:cs_pred_data}
\end{figure}

\vskip-1em
\noindent
We note that this model does not provide high predictive power in this case, with both algorithms finding parameter MLEs at the boundary of the search grid. We reiterate that the goal of this study is not to produce a high-quality model for this particular dataset, rather to employ the competing algorithms in a typical spatial estimation task for a realistic comparison of relative performance, and to demonstrate the low cost of surface prediction having performed an estimation task with FAIR.

\section*{Discussion}\label{sec:discussion}
\par In this work, we propose a DFT based approach to evaluate integrals that  are intrinsic to spatial prediction and inference when observations are integrals over irregular regions.  We show that for a stationary covariance function and correlation ranges that are a modest fraction of the spatial domain, the FAIR algorithm can be an order of magnitude or more faster than a direct quadrature approach. Moreover the FAIR algorithm provides, as an intermediate quantity, the matrix multiplication needed for spatial prediction on a grid. In either case the efficiency is due to replacing the double sum over grid points in the direct quadrature with a single sum and an FFT.  In addition the FAIR algorithm can reduce the number of evaluations of the covariance function and therefore can avoid many extra  transcendental and Bessel function evaluations. The application of change of support models to larger spatial data sets has been  limited by the computational burden of using direct quadrature. We believe the speedup afforded by FAIR algorithm now makes  it feasible to tackle larger problems. \\

\vskip-1em
\noindent One disadvantage of this algorithm are in cases when the correlation is large. Often a low order polynomial in the spatial coordinates, termed a spatial drift, is a useful component that provides an interpretable base model and also reduces large scale correlation.  Adding a low order polynomial to the spatial model can often reduce the correlation range in the stochastic component and so make the FAIR approximation accurate. This strategy raises the practical modeling question of whether large scale structure in a spatial structure is best represented as a low dimensional basis expansion or a covariance function with long range dependence. Typically either model will be effective and the low dimensional basis has the advantage of supporting the FAIR algorithm and often simplifying the dependence structure. \\

\vskip-1em
\noindent When the number of regions is moderate it is efficient to compute the DFTs of the the regions,  $\widehat{\bbB} ^i = DFT(\mathcal{I}_{B_i }) $ once and store the the results. For large grids and many regions this may not be possible due to limitation in memory. 
A two-dimensional DFT at a resolution of $2^{10}\times 2^{10}$ has a memory footprint of 16Mb (double complex precision).  A standard computational node ( for example a node on the NCAR supercomputer Cheyenne)  will have no difficulty storing several hundred regions, but will be unable to manage this task across, for example, $10,000$ regions. The alternative approach is to compute the transforms in smaller batches and recompute  as needed. Of course there is also the shortcut of considering single precision and a real-valued DFT to further reduce storage.  We believe this will still be more efficient than a direct quadrature approach noting that for a large number of regions the direct approach will involve many additional evaluations of the covariance function in the double sum.  Finally,  for regions that are translations of a single shape, such as the regular footprints from remote sensing platforms, one may be able to obtain $\widehat{\bbB} ^i$ directly using the fact that the translation operator is just a multiplication by complex exponentials in the transform space. \\

\vskip-1em
\noindent  Large spatial domains tend to exhibit  heterogeneity due to changing processes and conditions and so it is natural to consider nonstationary covariances to approximate variation in the process.  In this case we propose to extend the FAIR strategy to a fixed rank approach that uses many compactly supported basis functions and a sparse precision matrix for the basis coefficients (e.g., LatticeKrig model proposed by \cite{nychka:2015}) .  Explicitly, if  
\[ Y(\bbs) = \sum_j \psi_j(\bbs) c_j, \]
where $\{ \psi_j \}$ are a basis and $\bbc$ multivariate Gaussian 
then  a mean zero  observational model is 
 \begin{equation}
 \label{LKrig}
  \bbZ_i =  \int_{B_i} Y(\bbs) ds + \beps_i =  \sum_j  X_{i,j} c_j +\beps_i 
  \end{equation}
 with 
 \[ X_{i,j}= \int_{B_i}  \psi_j(\bbs) ds. \]
 
\noindent If the basis functions are built from translations of a single template then computing $X_{i,j}$ can be rephrased as a convolution of the region, $i$ with the template  translated to location, $j$ and computed efficiently  by modifying the FAIR algorithm. We note that the LatticeKrig \texttt{R} package supports nonstationary covariances and already implements the model given in Eqn.~(\ref{LKrig}) provided $\bbc$ follows a Gaussian Markov range field, $\bbX$  is sparse, and $\bbX$ is computed externally to the package code.   However, implementing nonstationary covariance models is very much an active area of research and the extension to change of support observations is challenging. \\

\vskip-1em
\noindent In summary,  providing an efficient algorithm for handling large change of support problems
will support spatial analysis of many new remotely-sensed and demographic data sets. We also hope this will open up more foundational research as to the value and potential limitations of considering aggregated spatial observations. 

\section*{Acknowledgement}
This research supported in part by Colorado School of Mines, faculty development funds, and NSF awards DMS-1854181 and DMS-1811384.

\bibliographystyle{plain}  

\bibliography{references}

\end{document}